\documentclass[aps,twocolumn,showpacs]{revtex4}
\pdfoutput=1
\makeatletter

\newcommand{\Rmnum}[1]{\expandafter\@slowromancap\romannumeral #1@}
\makeatother
\usepackage{graphicx}% Include figure files
\usepackage{dcolumn}% Align table columns on decimal point
\usepackage{bm}% bold math
\usepackage{CJK}
\usepackage{amsfonts}
\usepackage{psfrag}
\usepackage{wrapfig}
\usepackage{subfigure}
\usepackage{makeidx}
\usepackage{multirow}
\usepackage{epsf}
\usepackage{hyperref}
\usepackage{amsmath}%»¨Ìå×Öĸ
\usepackage{cases}
\DeclareMathOperator{\sech}{sech}

\begin{document}
\title{Interference properties of two-component matter wave solitons}
\author{Yan-Hong Qin$^{1,2}$}
\author{Yong Wu$^3$}
\author{Li-Chen Zhao$^{1,2}$}\email{zhaolichen3@nwu.edu.cn}
\author{Zhan-Ying Yang$^{1,2}$}
\address{$^{1}$School of Physics, Northwest University, Xi'an 710127, China}
\address{$^{2}$Shaanxi Key Laboratory for Theoretical Physics Frontiers, Xi'an 710127, China}
\address{$^{3}$School of Public Management, Northwest University, Xi'an 710127 China}

 %%%%%%%%%%%%%%%%%%%%%%%%%%%%%%%%%%%%%%%%%%%%%%%%%
\date{\today}
\begin{abstract}
The wave properties of solitons in a two-component Bose-Einstein condensates with attractive interactions or repulsive interactions are investigated in detail. We demonstrate that dark solitons in one of component admit interference and tunneling behaviour, in sharp contrast to the scalar dark solitons and vector dark solitons. The analytic analysis of interference properties shows that spatial interference pattern is determined by the relative velocity of solitons, while temporal interference pattern depends on the velocities and widths of two solitons, differing from the interference properties of scalar bright solitons. Especially, for attractive interactions system, we show that interference effects can induce some short-time density humps (whose densities are higher than background density) during the collision process of dark solitons. Moreover, the maximum hump value is remarkably sensitive to the variation of the solitons' parameters. For repulsive interactions system, the temporal-spatial interference periods have lower limits. Numerical simulations results suggest that interference patterns for dark-bright solitons are more robust against noises than bright-dark solitons. These explicit interference properties can be used to measure the velocities and widths of solitons. It is expected that these interference behaviour can be observed experimentally and could be used to design matter wave soliton interferometry in vector systems.

\end{abstract}

\maketitle

\section{INTRODUCTION}

Bose-Einstein condensate (BEC) is a prototypical quantum many-body systems. In the framework of mean-field theory, the dynamics of BECs systems are commonly described by nonlinear Schr\"{o}dinger equation (NLSE), also known as the Gross-Pitaevskii (GP) equation \cite{GPE}. The atomic interactions are described by a nonlinear term proportional to the s-wave scattering length and the condensates density. Therefore BECs provide a good platform to study solitons excitations \cite{BEC1,BEC2}. Bright solitons \cite{Bs1,Bs2,Bs3,Bs4} and dark solitons \cite{Ds1,Ds2,Ds3,Ds4} are observed in BECs with attractive and repulsive interatomic interactions, respectively. Solitons admit both particle and wave properties. The interactions between solitons are usually elastic just like particles \cite{particle1,particle2,particle3}. Recently, wave properties of solitons were discussed intensely \cite{zhaoguo,bsinter1,bsinter2,bsinter3,bsinter4,bsinter5,nodyzhao1,Nguyen,nodyzhao2,tunneling2}, mainly including interference behaviour and tunneling dynamics, since wave properties can be used to design matter wave soliton interferometry with high precision \cite{bsinter1,bsinter2,bsinter3,bsinter4,bsinter5}. With respect to the interference behaviour, bright soliton interferometry was proposed in BECs with attractive interactions \cite{bsinter1,bsinter2,bsinter3,bsinter4,bsinter5}. The interference and tunneling properties of scalar bright solitons have been described analytically in BECs \cite{nodyzhao1,nodyzhao2}. In contrary to scalar bright solitons, scalar dark solitons do not admit interference or tunneling behavior. However, dark soliton is another common soliton excitation, which can be used to measure physical quantities in BECs with repulsive interactions. Therefore, we would like to discuss the wave properties of vector solitons related to dark soliton in multi-component BECs.

The multi-component BECs, far from being a trivial extension of the single-component one, have shown many novel and fundamentally different dynamical behaviours \cite{MBEC1,MBEC2,MBEC3,MBEC4,MBEC5}. Recently, it was shown that tunneling oscillations could be observed between the interaction of two dark solitons in binary repulsive BECs \cite{tunneling2}. It indicated that dark solitons in one of component can admit wave properties when it is coupled with bright solitons in the other component. Then, it is natural to expect that dark solitons can allow interference behaviour in multi-component BECs, and vector solitons can be used to measure more physical parameters than scalar solitons. Therefore, we intend to study interference properties of bright-dark solitons as well as dark-bright solitons in a two-component BECs system. Similar discussions can be extended to more components coupled BECs.

In this paper, we mainly study the interference properties of bright-dark solitons (dark-bright solitons) in a two-component BECs with attractive (repulsive) interactions. We show that dark solitons in one of component can admit interference and tunneling behavior due to the feedback of the wave properties of bright-soliton component onto the dark one, in sharp contrast to the scalar dark solitons and dark-dark solitons. The explicit interference periods are characterized analytically, which suggests that interference patterns could be manipulated precisely by controlling the velocities and widths of two solitons. For attractive interactions, particularly, we note that the collision of dark solitons can induce some short-time humps above the background density in dark-soliton component. The detailed analyses show that the maximum hump value of dark-soliton component is sensitive to the relative phase, relative velocity, and relative width of two solitons. Additionally, we exhibit tunneling dynamics of solitons in both components. For repulsive interactions, temporal and spatial interference periods are found to both have lower limits. The maximum density value of dark-soliton component is equal to the background, differing substantially from the attractive interactions system.

The remainder of the present paper is organized as follows. In Sec.\Rmnum{2}, we introduce the theoretical model and present the bright-dark solitons solutions. In Sec. \Rmnum{3}, we analyze detailedly the wave properties of bright-dark two solitons during the interactions process in BECs with attractive interactions, mainly including interference patterns, short-time humps in dark-soliton component, and tunneling dynamics. In Sec. \Rmnum{4}, we explore the interference behaviour of dark-bright solitons in BECs with repulsive interactions. The analysis shows that the temporal and spatial interference periods have lower limits. Additionally, we demonstrate the stability of dark-bright solitons by numerical simulations. Finally, we summarize our results in Sec. \Rmnum{5}.

\section{Theoretical model and bright-dark soliton solutions}

The mean-field dynamics of a two-component BECs with attractive interactions is governed by the following equations \cite{GuoLing,lingzhao1}:
\begin{equation}\label{two-mode1}
    \begin{split}
      {\rm i}q_{b,t}+\frac{1}{2}q_{b,xx}+(|q_b|^2+|q_d|^2)q_b&=0, \\
      {\rm i}q_{d,t}+\frac{1}{2}q_{d,xx}+(|q_b|^2+|q_d|^2)q_d&=0.
    \end{split}
\end{equation}
where $q_{b}(x,t)$ and $q_{d}(x,t)$ represent the mean-field wave functions of bright-soliton component and dark-soliton component respectively in two-component BECs with attractive interactions. The well-known bright-dark soliton have been obtained in \cite{BD1,BD2,BD3,BD4}. To study the collision dynamics of bright-dark solitons, we re-derive solitons solutions by performing Darboux transformation \cite{lingzhao1} with the seed solutions $q_{0b}=0$ and $q_{0d}=e^{it}$. For simplicity and without lossing of generality, we discuss the collision dynamics of two bright-dark soliton based on exact solutions. The two-soliton solutions are
\begin{equation}\label{two-solution}
    \begin{split}
      q_{2b}&=q_{1b}-\frac{i2\lambda_{2i}\Phi^*_1\Phi_2}{|\Phi_1|^2+|\Phi_2|^2+|\Phi_3|^2}, \\
      q_{2d}&=q_{1d}-\frac{i2\lambda_{2i}\Phi^*_1\Phi_3}{|\Phi_1|^2+|\Phi_1|^2+|\Phi_3|^2}.
    \end{split}
\end{equation}
where $*$ means the complex conjugation. The explicit expressions for $\Phi_1,\Phi_2,\Phi_3$ are presented in Appendix. $q_{1b}$ and $q_{1d}$ are bright-dark one-soliton solutions. $q_{1b}=w_1\big[1+\frac{1}{w_1^2\!+\!v_1^2}\big]^\frac{1}{2}\sech[w_1(x\!-\!v_1t)-\eta/2]e^{i\beta_1+it},
q_{1d}=\frac{1}{v_1\!-\!iw_1}\big\{v_1\!-\!iw_1\tanh[w_1(x\!-\!v_1t)-\eta/2]\big\}e^{it},\eta=\ln(1+\frac{1}{v_1+w_1^2})$. Based on the bright-dark two-soliton solutions Eqs.\eqref{two-solution}, the asymptotic expressions of bright-soliton component $q_{2b}$ before the collision take the following forms (in the limit $t\rightarrow-\infty$ with assuming $v_1>v_2,w_1,w_2>0$):
\begin{equation}\label{asymptotic expression}
    \begin{split}
      BS_1&=c_1\sech[w_1(x-v_1t)+\frac{\eta_1}{2}]e^{i\beta_1+t}, \\
      BS_2&=c_2\sech[w_2(x-v_2t)+\frac{\eta_2}{2}]e^{i\beta_2+t}.
    \end{split}
\end{equation}
$BS_1$ and $BS_2$ correspond to the first bright soliton and the second bright soliton in component $q_{2b}$ before the collision, respectively. $\beta_j=i[v_jx+\frac{1}{2}(w_j^2-v_j^2)t-\varphi_j]$, $j=1,2$ and
\begin{equation}
\begin{split}
c_1&=-iw_1\frac{(1+v_1^2+w_1^2)}{(v_1^2+w_1^2)}\frac{(v_1-v_2+iw_1-iw_2)}{(v_1-v_2-iw_1-iw_2)},\nonumber\\
c_2&=-iw_2\frac{(1+v_2^2+w_2^2)}{(v_2^2+w_2^2)}\frac{(v_1-v_2-iw_1-iw_2)}{(v_1-v_2+iw_1-iw_2)}\\
&\times\frac{(v_1+iw_1)[1+(v_1-iw_1)(v_2+iw_2)]}{(v_1-iw_1)[1+(v_1+iw_1)(v_2+iw_2)]},\nonumber\\
e^{\eta_1}&=\frac{(v_1^2+w_1^2)}{(1+v_1^2+w_1^2)}\frac{(v_1-v_2)^2+(w_1+w_2)^2}{(v_1-v_2)^2+(w_1-w_2)^2},\nonumber\\
e^{\eta_2}&=\frac{(v_2^2+w_2^2)}{(1+v_2^2+w_2^2)}\frac{(v_1-v_2)^2+(w_1-w_2)^2}{(v_1-v_2)^2+(w_1+w_2)^2}\\
&\times\frac{(v_1v_2\!+\!1)^2\!+\!(w_1w_2\!-\!1)^2\!+\!v_2^2w_1^2\!+\!v_1^2w_2^2\!-\!1}{(v_1v_2\!+\!1)^2+(w_1w_2\!+\!1)^2\!+\!v_2^2w_1^2\!+\!v_1^2w_2^2\!-\!1}.\nonumber\\
\end{split}
\end{equation}
In above expressions, the parameters $v_j$ and $w_j$ correspond to two solitons' velocities and widths respectively. $\varphi_j$ is the initial phase of two bright solitons. Based on the asymptotic expressions \eqref{asymptotic expression}, the peak values of two bright solitons can be calculated as $P_j=w_j\big[\frac{1+v_j^2+w_j^2}{v_j^2+w_j^2}\big]^{\frac{1}{2}}$. It is seen that the peak values of two bright solitons in the component $q_{2b}$ are depend on both the widths and velocities of two solitons. Namely, the amplitude is no longer a independent physical parameter. This is different from the case for scalar bright solitons \cite{nodyzhao1}, for which the soliton amplitude does not depend on the moving velocity. Therefore, the velocity and width are two independent physical parameters for bright-dark soliton.

\section{Bright-dark solitons collision}

We study the collision process of two solitons based on the two-soliton solutions Eqs.\eqref{two-solution}. There are mainly three striking characters: interference patterns, humps induced by dark solitons interactions, and tunneling behavior, in contrast to the scalar dark solitons and dark-dark solitons \cite{Ds1,Ds2,Ds3,Ds4}. In what follows, we discuss them separately.

\begin{center}
A. Interference pattern
\end{center}

\begin{figure}[htbp]
\begin{center}
\subfigure{\includegraphics[width=85mm]{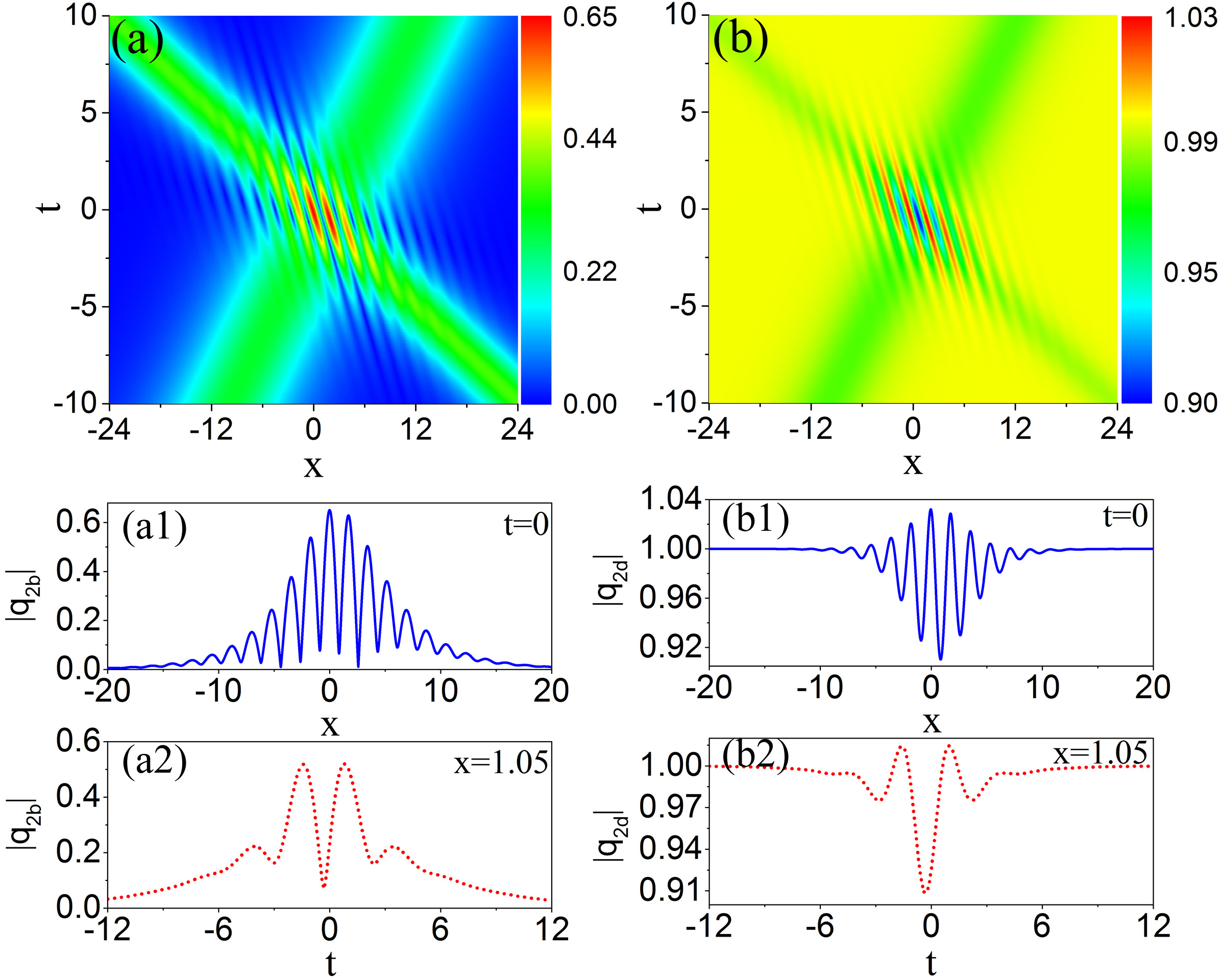}}
\end{center}
\caption{Interference patterns between bright-dark two solitons. Top panel: density evolution for solitons' collision, (a) for component $q_{2b}$ and (b) for component $q_{2d}$, respectively. Bottom panel: spatial interference fringes (blue solid line) and temporal interference fringes (red dotted line); They are the cutaway view of the interference patterns in Figs.\ref{Fig1}(a) and \ref{Fig1}(b) at t=0 and x=1.05 respectively. (a1) and (a2) correspond to component $q_{2b}$, (b1) and (b2) correspond to component $q_{2d}$. It is shown that interference patterns are formed in both components. Moreover, interference process involves two periods, one in the spatial direction and the other in the temporal direction. For dark solitons, interference behaviour was absent in the previous study. The parameters are $v_1=-2.4, v_2=1.1, w_1=0.324, w_2=0.224, \varphi_1=0,\varphi_2=0$.}\label{Fig1}
\end{figure}

As we know, solitons admit both particle and wave properties. The interference behaviour is a remarkable characteristic of the wave properties of solitons. The interference properties of scalar bright solitons and bright solitonic matter-wave interferometer have been widely investigated in nonlinear systems \cite{bsinter1,bsinter2,bsinter3,bsinter4,bsinter5,nodyzhao1,interfer1,interfer2}. However, neither scalar dark solitons nor vector dark solitons admits interference behaviour. Interestingly, during the collision process of bright-dark two solitons, not only the collision between two bright solitons generates the interference pattern in component $q_{2b}$, but two dark solitons' interplay can also yield the interference pattern in component $q_{2d}$. As an example, we show one case in Fig.\ref{Fig1} by choosing parameters $v_1=-2.4, v_2=1.1, w_1=0.324, w_2=0.224, \varphi_1=0,\varphi_2=0$. The top panels show the density distributions of temporal-spatial interference patterns; Figs. \ref{Fig1}(a) and \ref{Fig1}(b) correspond to the component $q_{2b}$ and the component $q_{2d}$, respectively. The bottom panels depict spatial interference fringes (at t=0, blue solid line) and temporal interference fringes (at x=1.05, red dotted line). These figures clearly demonstrate the interference behaviour of bright-dark solitons. It is seen that the temporal-spatial interference pattern shown in the component $q_{2b}$ gives excellent agreement with scalar scenario in Ref.\cite{nodyzhao1} [see Figs.\ref{Fig1}(a), \ref{Fig1}(a1) and \ref{Fig1}(a2)]. However, interference behaviour is very unusual for dark solitons in component $q_{2d}$ [see Figs.\ref{Fig1}(b), \ref{Fig1}(b1) and \ref{Fig1}(b2)], because scalar dark soliton and vector dark solitons do not admit wave properties due to its effective negative mass nature. This indicates that the nonlinear interaction between two components makes the interference behaviour in component $q_{2b}$ induce two dark solitons' collision to generate the interference pattern in component $q_{2d}$ simultaneously. Namely, due to the feedback of the wave properties of bright soliton onto the dark one, dark solitons can interfere with each other.

It is important to emphasize that the interference patterns can not always be observed during solitons' interactions processes, since the interference periods should be smaller than soliton scales for visible interference fringes \cite{nodyzhao1,nodyzhao2}. By means of the asymptotic analysis technic, the spatial and temporal periods are calculated as
\begin{align}
S=&\frac{2\pi}{|v_1-v_2|} \label{D} \\
T=&\frac{4\pi}{|v_2^2-v_1^2+w_1^2-w_2^2|} \label{T}
\end{align}
\emph{``S''} and \emph{``T''} denote spatial period and temporal period, respectively. Obviously, the relative velocity ($v_1-v_2$) determines the spatial interference pattern Eq.(\ref{D}), while temporal period is determined by both widths $w_j$ and velocities $v_j$ of two solitons Eq.(\ref{T}). This is quite different from the interference properties of the scalar bright soliton reported previously in Ref.\cite{nodyzhao1}, in which temporal period was depended on both the peaks and velocities. By applying the spatial-temporal period expressions Eqs.(\ref{D}-\ref{T}), the interference patterns can be manipulated by controlling solitons' velocities and widths. When the absolute values of velocities of two solitons are identical, the spatial interference pattern will not be formed; When two solitons are of the same width and equal velocity squared, the temporal pattern will disappear. Based on the matter wavelength theory \cite{matter wave}, the temporal-spatial interference patterns are visible when the soliton parameters satisfy the condition that spatial period $S$ is smaller than the scales of two solitons and temporal period $T$ is smaller than the time scale of collision (as shown in Fig.\ref{Fig1}).

Particularly, we note that two dark solitons' interaction form some short-time humps above the background density by their interference effects, as shown in the right panel of Fig.\ref{Fig1}, in sharp contrast to scalar dark solitons and dark-dark solitons. This clearly indicates that bright solitons could induce dark solitons to allow much richer dynamics than their own in the repulsive interaction systems. This point is further investigated in the following text.

\begin{center}
B. The maximum hump density value in the dark-soliton component
\end{center}

\begin{figure}[htbp]
\begin{center}
\subfigure{\includegraphics[width=85mm]{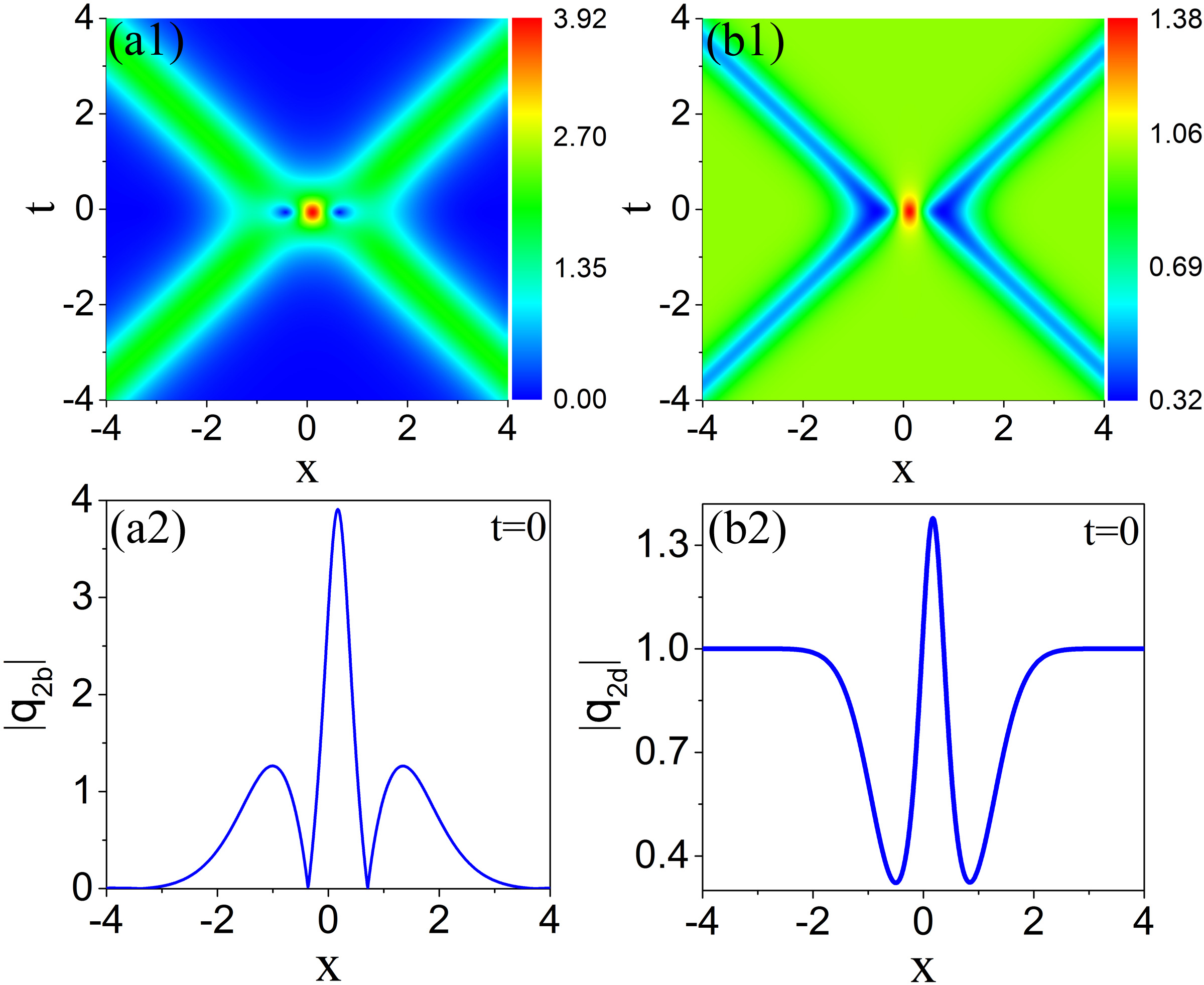}}
\end{center}
\caption{The interaction of bright-dark two solitons. Top panel: density distributions of two solitons' collision in two components; (a1) for component $q_{2b}$ and (b1) for component $q_{2d}$. Bottom panel: intensity profiles for both components at t=0; (a2) and (b2) correspond to the component $q_{2b}$ (thin blue line) and the component $q_{2d}$ (thick blue line) respectively. It is seen that two dark solitons' collision induces a very high density hump above the background density in component $q_{2d}$. The parameters are $w_1=1.8,v_1=-1,w_2=1.8,v_2=1,\varphi_1=1.835\pi,\varphi_2=0$.}\label{Fig2}
\end{figure}

In general, dark solitons collide elastically and could produce some dips in collision region in the repulsive interaction system \cite{lingzhao2}. Nevertheless, we have shown that dark solitons' collision can form some short-time density humps above the background density in the attractive interaction system Eqs.\eqref{two-mode1}, as depicted in the right panel of Fig.\ref{Fig1}. To show this character more clearly, we plot Fig.\ref{Fig2} by setting parameters $w_1=1.8,v_1=-1,w_2=1.8,v_2=1,\varphi_1=1.8\pi,\varphi_2=0$. Figs. \ref{Fig2}(a1) and \ref{Fig2}(b1) correspond to the density evolutions of component $q_{2b}$ and component $q_{2d}$, respectively. A highlighted feature is that the interactions of two solitons in both components generate a high hump and two valleys in the collision center simultaneously. The corresponding intensity profiles at t=0 (shown in Figs.\ref{Fig2}(a2) and \ref{Fig2}(b2)) describe this characteristic more evidently. Of particular note is that the hump appearing in component $q_{2d}$ is significantly higher than the background density. This dynamical behaviour is also not admitted for scalar dark solitons and vector dark solitons. It should be mentioned that the humps produced by the dark solitons' interaction are not always visible. One has to wonder how does the soliton parameters $w_j,v_j,\varphi_j$ affect the hump values in component $q_{2d}$? For simplicity, we choose Fig.\ref{Fig2} as an example to discuss the changes of maximum hump value of component $q_{2d}$ at t=0, with varying the relative phase, relative velocity and relative width of two solitons by means of the control variate method.

\begin{figure}[htbp]
\centering
\subfigure{\includegraphics[width=86mm]{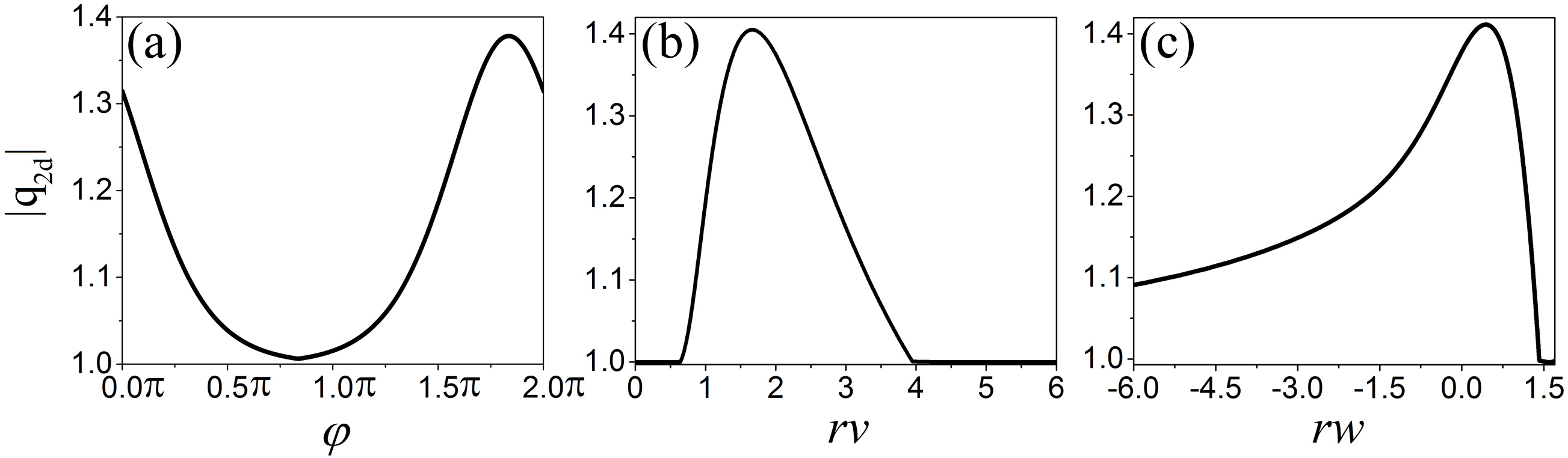}}
\caption{The variation of maximum hump value in dark-soliton component $q_{2d}$ vs the solitons' parameters. (a) The maximum hump value varies with the relative phase ($\varphi$) for the choice of parameters $w_1=1.8,v_1=-1,w_2=1.8,v_2=1,\varphi_1=\varphi,\varphi_2=0$. (b) The maximum hump value varies with the relative velocity (rv) between two solitons, with parameters setting $w_1=1.8,v_1=v_2-rv,w_2=1.8,v_2=1,\varphi_1=1.835\pi,\varphi_2=0$. (c) The maximum hump value varies with the relative width (rw) between two solitons with parameters $w_1=w_2-rw,v_1=-1,w_2=1.8,v_2=1,\varphi_1=1.835\pi,\varphi_2=0$. It is shown that the maximum density hump value is sensitive to soliton parameters.}\label{Fig3}
\end{figure}

Firstly, we study the effect of the relative phase between bright solitons on the maximum density value at t=0 in component $q_{2d}$. Generally, in two-soliton circumstance, one soliton can be regarded as the reference ($\varphi_2=0$ herein), and the other soliton's initial phase ($\varphi_1$) will become the relative phase (denoted by $\varphi$) between two bright solitons. Then, one can observe the change of the maximum hump value of component $q_{2d}$ as the variation of the $\varphi$ value. As presented in Fig.\ref{Fig3}(a), $\varphi=\varphi_1$ and the other parameters are identical with the ones in Fig.\ref{Fig2}. Obviously, the hump value $|q_{2d}|$ is very sensitive to the relative phase $\varphi$ of two bright solitons. With the increasing of $\varphi$, the hump value of component $q_{2d}$ appears decreasing at first and then increasing gradually after reaching the minimum hump value (at $\varphi\approx0.835\pi$) which approximates the background density. Subsequently, the hump value reaches to the maximum value (at $\varphi\approx1.835\pi$). It indicates that the density distribution of component $q_{2d}$ strongly depends on the relative phase between two bright solitons in component $q_{2b}$. This character could be used to measure the relative phase between solitons.

In the second scenario, we investigate the changes of the maximum hump value of component $q_{2d}$ at t=0 by varying the relative velocity ($rv$) of two solitons by setting the parameter $v_1=v_2-rv$, and the other parameters are identical with the ones in Fig.\ref{Fig2}. As shown in Fig.\ref{Fig3}(b), there are two critical relative velocities, around $rv=0.63$ and $rv=3.91$ respectively. When $rv\leq0.63$ or $rv\geq3.91$, the change of the maximum density value is nearly imperceptible with the increasing of rv value. When $0.63<rv<3.91$, the density value firstly increases to the maximum value at $rv\approx1.66$ and then decreases to be very closer to the background density with the increasing of the $rv$ value. The critical values of relative velocity vary with the relative phase between bright solitons. The underlying definite properties still need further studies.

Thirdly, we investigate the change of the maximum hump value of component $q_{2d}$ at t=0 by varying the relative width ($rw$) between two solitons by setting the parameters \rm$w_1=w_2-rw$ and the other parameters are identical with the ones in Fig.\ref{Fig2}. This is depicted in Fig.\ref{Fig3}(c). Note that the maximum density value becomes progressively discernible as the increasing of $rw$ value. When the difference between their width value keeps decreasing, the maximum density value continues to increase. When $rw\approx0.44$, the density of component $q_{2b}$ reachs the highest value. With the further increasing of $rw$ value, the width of the first soliton tend to be very small, the maximum density value decrease rapidly being close to the background density. This reveals that the scales of two solitons dramatically affect the density distribution of component $q_{2d}$. It should be pointed out that when the relatively velocity of two solitons is relatively large, the change of the $rw$ value has no significant effect on the maximum density value of component $q_{2d}$.

More recently, a method was proposed to split the ground state of an attractively interacting BEC into two bright solitary waves with controlled relative phase and velocity \cite{RP}. Combining above discussions, one expects these properties could be used to test some physical quantities related to solitons in the near-future experiments.

\begin{center}
C. Tunneling behaviour
\end{center}
\begin{figure}[htbp]
\begin{center}
\subfigure{\includegraphics[width=85mm]{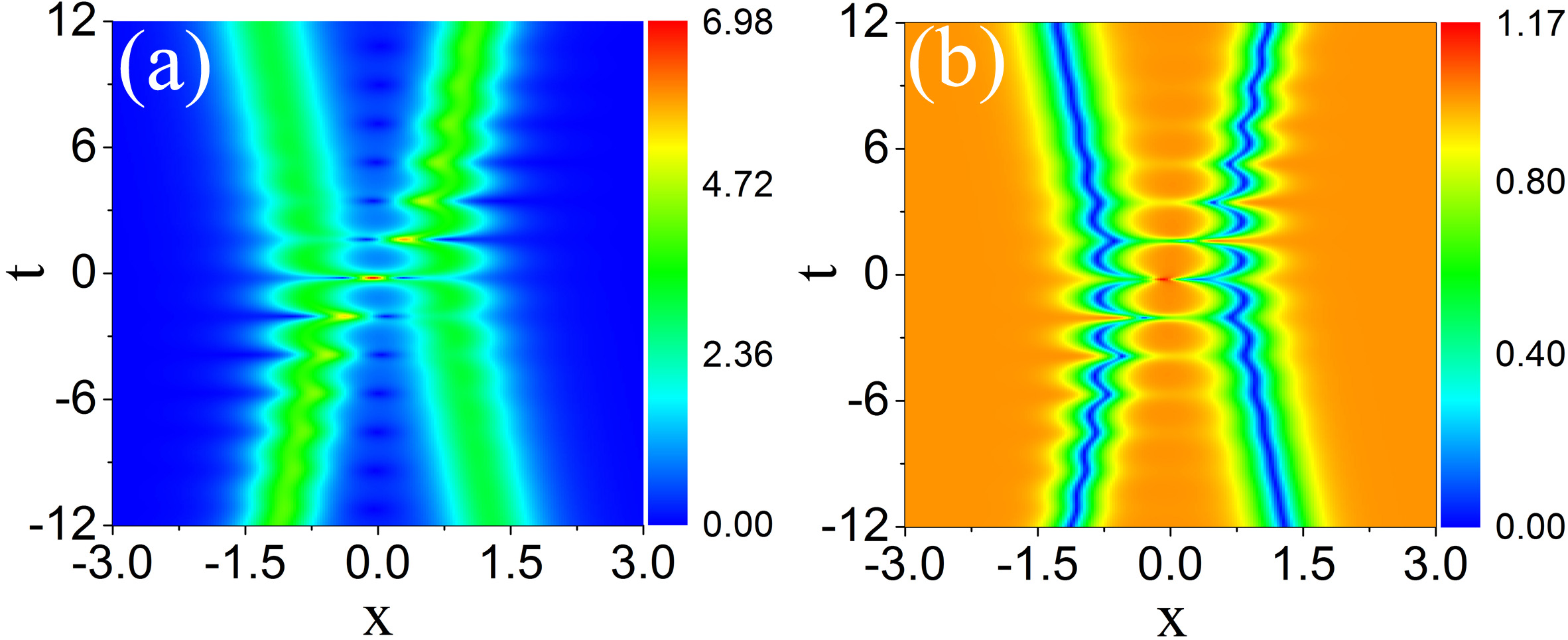}}
\caption{The tunneling behaviour between two bright-dark solitons. (a) shows the density of the bright-soliton component, (b) for the dark-soliton component. It is seen that the oscillating tunnleing behavior emerges during the interaction process. Parameters are $w_1=2.9,v_1=-0.05,w_2=3.9,v_2=0.05,\varphi_1=\varphi_2=0$.}\label{Fig4}
\end{center}
\end{figure}

The quantum tunneling dynamics of solitons have been discussed well in \cite{tunneling1,tunneling2,tunneling3,tunneling4,tunneling5,tunneling6,tunneling7,nodyzhao2}. Recently, tunneling dynamics of dark solitons in a harmonic trap were investigated in binary repulsive BECs \cite{tunneling2}, based on different initial conditions of the phase difference and population imbalance of bright solitons. Then, it would be natural to expect that the tunneling dynamics between dark solitons can also be observed in the coupled system Eqs. \eqref{two-mode1} .

\begin{figure}[htbp]
\begin{center}
\subfigure{\includegraphics[width=85mm]{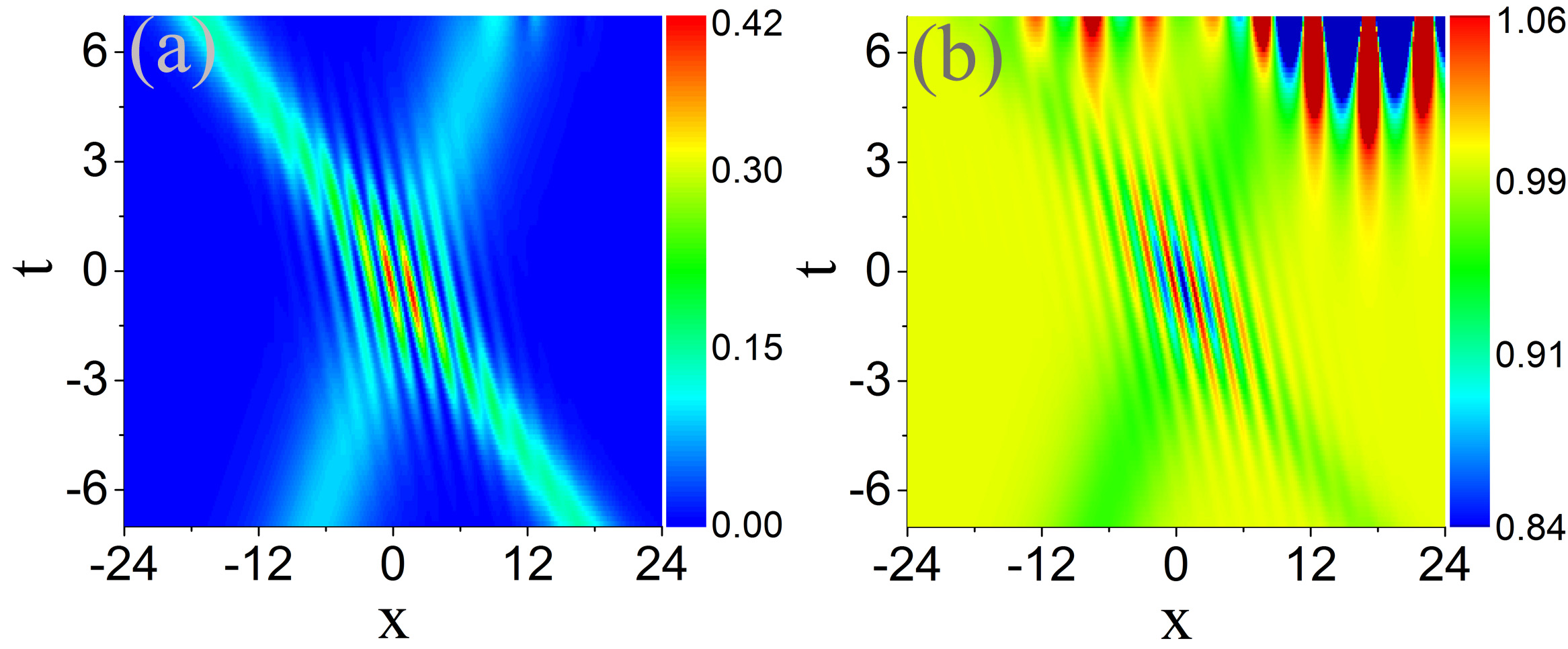}}
\end{center}
\caption{Numerical simulations of the bright-dark solitons with the same parameters as Fig.\ref{Fig1}, (a) for bright-soliton component $q_{2b}$, and (b) for dark-soliton component $q_{2d}$. It is seen that the background density presents unstable in dark-soliton component $q_{2d}$ after the time $t\approx1.5$.}\label{Fig5}
\end{figure}

Based on the quantum tunneling theory \cite{TT}, the nonlinear term $-(|q_1|^2+|q_2|^2)$ in Eqs.\eqref{two-mode1} can be seen as an effective double-well potential, which is self-induced by the distribution of atoms. The structure of quantum wells evolve simultaneously with the evolution of bright solitons and dark solitons in both components, since atoms tunneling from one soliton to the other change the quantum well structure synchronously. Therefore, we call it as the tunneling behavior of matter-wave solitons in a self-induced quantum wells, in contrast to the external double-well potential in usual quantum theory. One of the typical examples of the density for tunneling behaviour has been depicted in Fig.\ref{Fig4} by setting the parameters $w_1=2.9,v_1=-0.05,w_2=3.9,v_2=0.05,\varphi_1=\varphi_2=0$, (a) and (b) correspond to component $q_{2b}$ and component $q_{2d}$ respectively. It is seen that coupled nonlinear effects between two components force dark solitons performing periodic oscillation in time evolution together with bright solitons. In contrary, scalar dark solitons and dark-dark solitons do not allow these features due to its effective negative mass nature. The tunneling period is calculated as $T=4\pi/(v_2^2-v_1^2+w_1^2-w_2^2)$, determined by widths and velocities of solitons. For visible tunneling behaviour, the tunneling period should be smaller than the half of time scale of collision.  Tunneling dynamics shown in Fig.\ref{Fig4} is different from ones observed in \cite{tunneling2}, in which oscillations were related with the deviation from the in-phase or out-of-phase stationary solution.

Next, we discuss the stability of bright-dark solitons by numerical calculation. The numerical evolution results of bright-dark solitons are displayed in Fig.\ref{Fig5}, which initial excitation forms are given by the same parameters of Fig.\ref{Fig1} at $t=-7$, (a) for component $q_{2b}$ and (b) for component $q_{2d}$. It is seen that the numerical results in Fig.\ref{Fig5} reproduce the the interference fringes with high visibility when solitons collide with each other in both components, which agrees pretty well with the analytical results in Figs.\ref{Fig1}(a) and \ref{Fig1}(b). However, there are some other localized waves emerging in dark-soliton component when $t>1.5$, induced by the modulational instability of the background fields \cite{MI1}. In view of this fact, we would like to further investigate interference behavior of dark-bright solitons in two-component BECs with repulsive interactions, since the background field does not admit any modulational instability.

\begin{figure}[htbp]
\begin{center}
\subfigure{\includegraphics[width=85mm]{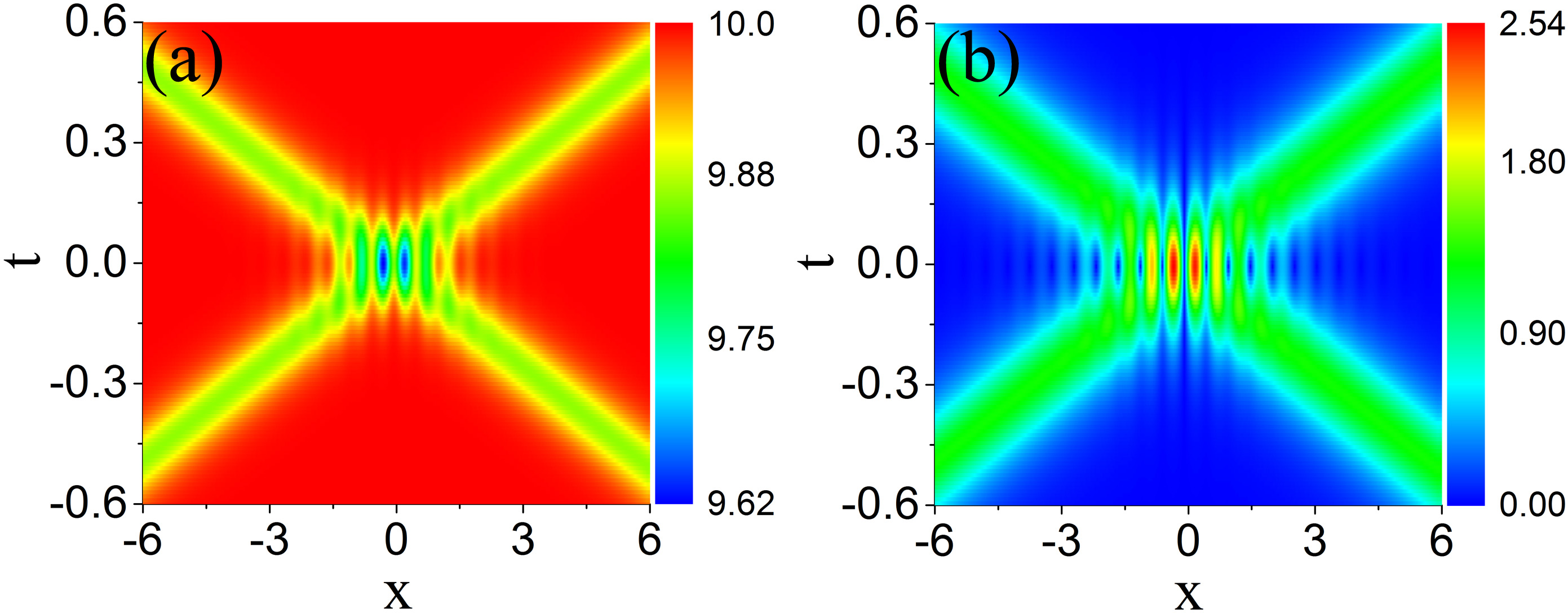}}
\caption{Interference patterns of two dark-bright solitons' interaction, (a) for dark-soliton component and (b) for bright-soliton component. The parameters are $\varrho_1=\varrho_2=1, \kappa_1=-\kappa_2=-6, q_0=10, x_1=x_2=0, \phi_1=\phi_2=\pi/4$.}\label{Fig6}
\end{center}
\end{figure}

\section{Dark-bright solitons collision}

For dark-bright solitons in two-component BECs with repulsive interactions, dark solitons in one component play the role of an effective potential that enables the bound-state trapping of the bright-soliton component \cite{Ds3,D-B1,D-B3,D-B4}. It is expected that the dark solitons can show more exotic dynamics behaviours in this system, when they are coupled with bright solitons. We consider the two-component CNLSE of the forms
\begin{equation}\label{two-mode2}
     \begin{split}
{\rm i}q_{d,t}+q_{d,xx}-2(|q_b|^2+|q_d|^2)q_d&=0, \\
{\rm i}q_{b,t}+q_{b,xx}-2(|q_b|^2+|q_d|^2)q_b&=0.
    \end{split}
\end{equation}
The dynamics of two dark-bright solitons can be described by the well-known exact solutions in Ref.\cite{D-B3}
\begin{align}
q_d&=q_0e^{-2iq_0^2t}\bigg\{1+\frac{1}{D}\bigg[\frac{\Gamma_1}{z_1(z_1^*-z_1)}+\frac{\Gamma_2}{z_2(z_2^*-z_2)}\nonumber\\
&+\Gamma_3\frac{e^{-i\beta}(q_0^2\!-\!z_1^*z_2)\delta_1^*\delta_2}{z_2(z_2\!-\!z_1^*)}+\Gamma_3\frac{e^{i\beta}(q_0^2-z_1z_2^*)\delta_1\delta_2^*}{z_1(z_1-z_2^*)}\nonumber\\
&+\frac{z_1^*z_2^*-z_1z_2}{z_1z_2}\Gamma_5\bigg]\bigg\},\label{d}\\
q_b&=-\frac{q_0e^{-2iq_0^2t}}{D}\bigg[\frac{\bar{\delta_1}z_1^*}{q_0^2}e^{-iz_1^*(x+z_1^*t)}+\frac{\bar{\delta_2}z_2^*}{q_0^2}e^{-iz_2^*(x+z_2^*t)}\nonumber\\
&+\Gamma_4\frac{\bar{\delta_1}\bar{\delta_2}z_1^*z_2^*(q_0^2-z_1^*z_2^*)(z_1^*-z_2^*)^2}{q_0^2(q_0^2-z_1z_2)}\bigg].\label{b}
\end{align}
\begin{figure}[htbp]
\begin{center}
\subfigure{\includegraphics[width=85mm]{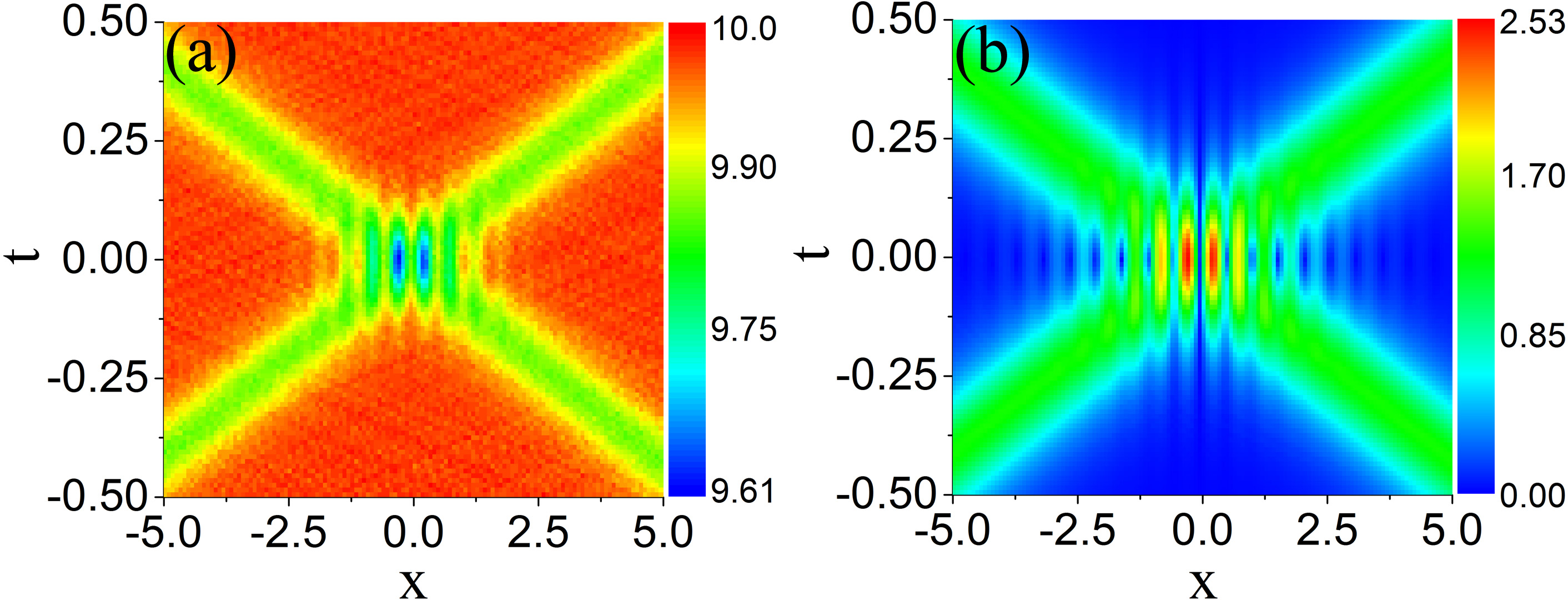}}
\caption{ The numerical simulations of the dark-bright solitons with small noise. (a) for dark-soliton component and (b) for bright-soliton component. The initial excitation condition is given by the same parameters as Fig.\ref{Fig6} at $t=-0.5$ by multiplying a factor $(1 + 0.02Random[-1,1])$. It is seen that the dark-bright solitons are robust against small noise.}\label{Fig7}
\end{center}
\end{figure}
where the explicit expressions for $D$, $\bar{\delta}_{1,2}$ and $\Gamma_i(i=1...5)$ are presented in Appendix. $q_d(x,t)$ and $q_b(x,t)$ describe the wave functions of dark-soliton component and bright-soliton component, respectively. $q_0$ is the amplitude of background. $z_j=\kappa_j+i\varrho_j$ corresponds to the eigenvalues of the inverse scattering transform problem with $|z_j|<q_0$, $j=1,2$. The soliton's velocity is $v_j=2\kappa_j$. $\delta_j=2\varrho_j/[q_0(q_0^2-z_j^2)^\frac{1}{2}]e^{x_j+i\phi_j}$ is so-called norming constants. The position offset and phase of solitons are described by $x_j$ and $\phi_j$ respectively.

Dark-bright solitons in repulsive interactions system Eqs.\eqref{two-mode1} have similar wave properties to bright-dark solitons in attractive interactions system Eqs.\eqref{two-mode2}, such as interference behaviour and tunneling dynamics. As an example, herein we show their interference patterns with high visibility in both components in Fig.\ref{Fig6}, (a) for component $q_{d}$ and (b) for component $q_{b}$ (see caption for details). By further simplifying solitons solutions \eqref{d}-\eqref{b} and combining the asymptotic analysis expressions in Ref.\cite{D-B3}, the interference periods can be obtained. The spatial interference period is $S=4\pi/|v_1-v_2|$, and temporal interference period is $T=2\pi/|w_1^2-w_2^2-\frac{1}{4}(v_1^2-v_2^2)|$, where $w_1$ and $w_2 $ are the widths of two solitons with $w_j=\varrho_j$. It is shown that the spatial interference period depends on the relative velocity of two solitons, and temporal interference period is determined by the velocities and widths of two solitons (herein the parameter settings in Fig.\ref{Fig6} make the temporal period be zero). Their interference properties are similar to the bright-dark solitons mentioned in Sec.III. But it must be emphasized that interference periods of dark-bright solitons are found to both have lower limits, namely, $S>\frac{\pi}{q_0}$ and $T>\frac{\pi}{q_0^2}$. This is induced by the velocity of the dark soliton is always lower than the sound speed of the system for the dark-bright soliton, which is different from the bright-dark soliton mentioned above. Moreover, we note that the maximum density value of dark-soliton component is always equal to background density, in stark contrast to bright-dark solitons in attractive interaction system, in which the collision of dark solitons can induce some short-time density humps by the interference effects (compare Fig.\ref{Fig1}(b) and Fig.\ref{Fig2}(b1) to Fig.\ref{Fig6}(a)).

We further perform numerical simulations to verify the stability of dark-bright solitons. Here, we simulate the initial excitation condition perturbed with $2\%$ white noise of small amplitudes. Namely, we multiply the $q_d$ and $q_b$ by the factors $(1 + 0.02Random[-1,1])$. For instance, we show the simulation results for two dark-bright solitons in Fig.\ref{Fig7}. The initial excitation forms are given by the exact ones at $t=-0.5$ in Fig.\ref{Fig6} with adding small noise. It is seen that, under the noise perturbation, simulation results reproduce the stable interference patterns in both components, which agree well with the analytical results Fig.\ref{Fig6}. It reveals that dark-bright solitons are very robust against perturbations than bright-dark solitons since there is no modulational instability for the coupled model with repulsive interactions. On the other hand, experimental observations demonstrated that two-component solitons can be produced based on well-developed density and phase modulation techniques \cite{Ds3,D-B4}. Solitons interactions have been also demonstrated experimentally in BECs \cite{RP,Nguyen}. Recently, interferometry with BECs in microgravity \cite{inter5}, spin-orbit-coupled interferometry \cite{inter6} and multicomponent interferometer in a spinor BECs \cite{inter7} were proposed, which demonstrate that the interference pattern holds great promise for implementing quantum tests and measurement information for uncorrelated systems. One can expect that the interference patterns and tunneling behaviour of vector solitons obtained here could be used to measure some physical quantities in some ultra-cold atomic gases.

\section{CONCLUSION AND DISCUSSION}

In this paper, we show that interactions between bright-dark two solitons and dark-bright two solitons both can generate temporal-spatial interference patterns which are not admitted for scalar dark solitons and vector dark solitons. The explicit interference periods expressions can be used to measure some physical quantities, such as soliton's velocity, width, and related acceleration fields. In attractive interactions system, the maximum density value of dark-soliton component can be higher than the background density and changes obviously as the variation of solitons' parameters. Additionally, we display the tunneling dynamics of bright-dark solitons. In repulsive interactions system, since solitons' velocity cannot exceed the speed of sound, the spatial-temporal interference periods both have lower limits. Moreover, we use numerical simulations to test the stabilities of the bright-dark solitons and dark-bright solitons. The results indicate that dark-bright solitons are more robust than bright-dark solitons as the background field does not admit any modulational instability for repulsive system. From a physical perspective, the nonlinear feedback of bright soliton into dark solitons leads the latter to admit more interesting dynamics in the other component, rather than only retaining the characters as scalar dark solitons or dark-dark solitons.

\section*{Acknowledgments}
This work is supported by National Natural Science Foundation of China (Contact No. 11775176), Basic Research Program of Natural Science of Shaanxi Province (Grant No. 2018KJXX-094), The Key Innovative Research Team of Quantum Many-Body Theory and Quantum Control in Shaanxi Province (Grant No. 2017KCT-12), and the Major Basic Research Program of Natural Science of Shaanxi Province (Grant No. 2017ZDJC-32).

\section*{Appendix}
The explicit expressions for $\Phi_1,\Phi_2,\Phi_3$ of Eqs.\eqref{two-solution} are
\begin{small}
\begin{equation}
\begin{split}
\Phi_1=&\frac{\lambda_2\Phi_{21}}{\lambda_2-\lambda^*_1}+\{\lambda_1\Phi_{11}(\Phi^*_{11}\Phi_{21}+\Phi^*_{12}\Phi_{22}+\Phi^*_{13}\Phi_{23})\\
&+\lambda^*_1[(|\Phi_{12}|^2+|\Phi_{13}|^2)\Phi_{21}\!-\!(\Phi^*_{12}\Phi_{22}+\Phi^*_{13}\Phi_{23})\Phi^*_{11}]\}/\\
&[(\lambda^*_1-\lambda_2)(|\Phi_{11}|^2+|\Phi_{12}|^2+|\Phi_{13}|^2)],\nonumber\\
\Phi_2=&\frac{\lambda_2\Phi_{22}}{\lambda_2-\lambda^*_1}+\{\lambda_1\Phi_{12}(\Phi^*_{11}\Phi_{21}+\Phi^*_{12}\Phi_{22}+\Phi^*_{13}\Phi_{23})\\
&+\lambda^*_1[(|\Phi_{11}|^2+|\Phi_{13}|^2)\Phi_{22}\!-\!(\Phi^*_{11}\Phi_{11}+\Phi^*_{13}\Phi_{23})\Phi^*_{12}]\}/\\
&[(\lambda^*_1-\lambda_2)(|\Phi_{11}|^2+|\Phi_{12}|^2+|\Phi_{13}|^2)],\nonumber\\
\Phi_3=&\frac{\lambda_2\Phi_{23}}{\lambda_2-\lambda^*_1}+\{\lambda_1\Phi_{13}(\Phi^*_{11}\Phi_{21}+\Phi^*_{12}\Phi_{22}+\Phi^*_{13}\Phi_{23})\\
&+\lambda^*_1[(|\Phi_{11}|^2+|\Phi_{12}|^2)\Phi_{23}\!-\!(\Phi^*_{11}\Phi_{21}+\Phi^*_{12}\Phi_{22})\Phi^*_{13}]\}/\\
&[(\lambda^*_1-\lambda_2)(|\Phi_{11}|^2+|\Phi_{12}|^2+|\Phi_{13}|^2)].\nonumber
\end{split}
\end{equation}
\end{small}
where $\Phi_{j1}=e^{\alpha_j-i\beta_j},\Phi_{j2}=e^{it},\Phi_{j3}=\frac{e^{\alpha_j-i\beta_j+it}}{iw_j-v_j},
\alpha_j=w_j(x-v_jt),\beta_j=v_jx-\frac{1}{2}(v_j^2-w_j^2)t-\varphi_j,\lambda_j=\lambda_{jr}+i\lambda_{ji},
\lambda_{ji}=w_j\big(\frac{1}{v_j^2+w_j^2}+1\big),\lambda_{jr}=v_j\big(\frac{1}{v_j^2+w_j^2}-1\big),j=1,2.$

The expressions for $\bar{\delta}_{1,2}$, $D$ and $\Gamma_i$ of Eqs.\eqref{d}-\eqref{b} are
\begin{small}
\begin{equation}
\begin{split}
\bar{\delta}_1&=-\frac{q_0^2(q_0^2-|z_1|^2)(q_0^2-z_1^*z_2)}{(z_1^*)^2(q_0^2-z_1^*z_2^*)}\delta_1^*,\nonumber\\
\bar{\delta}_2&=-\frac{q_0^2(q_0^2-|z_2|^2)(q_0^2-z_1z_2^*)}{(z_2^*)^2(q_0^2-z_1^*z_2^*)}\delta_2^*,\nonumber\\
\Gamma_1&=\frac{q_0^2(q_0^2-|z_1|^2)|q_0^2-z_1^*z_2|^2|\delta_1|^2}{|q_0^2-z_1z_2|^2}e^{-2\varrho_1\xi_1},\nonumber\\
\Gamma_2&=\frac{q_0^2(q_0^2-|z_2|^2)|q_0^2-z_1^*z_2|^2|\delta_2|^2}{|q_0^2-z_1z_2|^2}e^{-2\varrho_2\xi_2},\nonumber\\
\Gamma_3&=\frac{q_0^2(q_0^2-|z_1|^2)(q_0^2-|z_2|^2)}{|q_0^2-z_1z_2|^2}e^{-(\varrho_1\xi_1+\varrho_2\xi_2)},\nonumber\\
\Gamma_4&=\frac{\delta_1z_1e^{-2\varrho_1\xi_1-iz_2^*(x+z_2^*t)}}{(z_1^*-z_1)^2(z_1+z_2^*)^2}+\frac{\delta_2z_2e^{-2\varrho_2\xi_2-iz_1^*(x+z_1^*t)}}{(z_2^*-z_2)^2(z_2-z_1^*)^2}, \nonumber\\
\Gamma_5&=\frac{(q_0^2\!-\!|z_1|^2)(q_0^2\!-\!|z_2|^2)|q_0^2\!-\!z_1^*z_2|^2|z_1\!-\!z_2|^4}{16\varrho_1^2\varrho_2^2|q_0^2-z_1z_2|^2|z_1^*-z_2|^4}\nonumber\\
&\times q_0^4|\delta_1|^2|\delta_2|^2e^{-2(\varrho_1\xi_1+\varrho_2\xi_2)},\nonumber\\
D&=\Gamma_3\bigg[\frac{-e^{i\beta}(q_0^2-z_1z_2^*)\delta_1\delta_2^*}{(z_1-z_2^*)^2}\!-\!\frac{e^{-i\beta}(q_0^2-z_1^*z_2)\delta_1^*\delta_2}{(z_2-z_1^*)^2}\bigg]\nonumber\\
&+1+\frac{\Gamma_1}{4\varrho_1^2}+\frac{\Gamma_2}{4\varrho_2^2}+\Gamma_4.
\end{split}
\end{equation}
\end{small}
where $\xi_j=x+2\kappa_jt, j=1,2$

\end{document}